\documentstyle [12pt,openbib]{article}
\title{Gravitational field of an infinitely long supermassive cosmic string}
\author{M. M. Som \\
        Instituto de F\'{\i}sica\\Universidade Federal do Rio de
        Janeiro\\
        Rio de Janeiro, Brasil\\
        and\\
        Marcelo de Oliveira Souza\\
        Universidade Estadual do Norte Fluminense\\
        Centro de Ci\^encia e Tecnologia\\
        Campos, Rio de Janeiro, Brasil\\
        and\\
        Instituto de F\'{\i}sica\\Universidade Federal do Rio de
        Janeiro\\
        Rio de Janeiro, Brasil \\
       \bigskip
       PACS number(s): 04.20.Jb, 04.20.Cv \\
       \bigskip
       e-mail: som@if.ufrj.br and mm@uenf.br}

\date{11/11/94}
\begin{document}
\maketitle
\pagebreak
\begin{abstract}

We obtain an exact solution of the coupled Einstein-scalar-gauge
field equations for a local  infinitely long supermassive cosmic string.
The solution corresponds to that of Hiscok-Gott. The string appears
to be due to the freezing of the scalar field at the null value
giving rise to a constant linear energy density.

\end{abstract}
\pagebreak

Spontaneous symmetry breaking in gauge theories leads to
 phase transitions in the early Universe [1,2,3]. In this period can
appear topological structures in the Universe like strings,
vacuum domain walls and monopoles. One of these topological
defects that has drawn more
attention is the cosmic string. The cosmic strings are linear
topological defects. Only infinitely long and closed
loop strings can exist.
 An infinitely long cosmic string is a static cylindrically symmetric
configuration of self-interacting scalar field minimally coupled
to a U(1) gauge field.

We have two general type of strings. Those which arise in a phase
transition in wich a gauge symmetry undergoes spontaneous symmetry
breaking is called local or gauge strings, and those which we call
a global string, arise as a result of a spontaneous breaking
of global symmetry. Strings that have cosmological interest are
the local strings.

Local strings have a linear energy density $\mu \sim \eta^{2}$,
where $\eta$ is the vacuum expectation value (VEV) acquired
by the gauge field in the phase transition.

Strings of astrophysical relevance were formed during phase
transitions at the grand-unified-theory (GUT) scale ($10^{15}$ GeV).
They have linear energy density $\mu \sim \eta^{2}$ of the order
$10^{-6}$. This small value of $\mu$ could justify the weak-field
approximation used by Vilenkin.

Vilenkin [4] using the weak-field aproximation obtained
that the geometry around a cosmic string is conical, with
the deficit angle given by $\delta \phi = 8 \pi \mu$,
where $\mu$ is the linear energy density of the string.
Hiscock [5] and Gott [6] found independently an exact
spacetime metric corresponding to a static
cilindrically symmetric string, and showed that the
results obtained by Vilenkin are correct.
They utilized
a stress energy tensor  with
the energy density  constant within the string,
without considering the structure
of the scalar and the gauge fields.
 Raychaudhuri [7]
criticized these results giving a new interpretation to
their metric, using  the full equations, not
taken into account by them.
So far no solution in closed form is obtained of the
coupled Einstein-scalar-gauge field equations.

More massive strings, $\eta >> 10^{-3}$ could be produced,
because the Universe might have undergone phase transitions
at energy scales higher than the $10^{15}$ GeV GUT scale.
Due to their large masses, $\mu >> 10^{-6}$, we believe it
is not reasonable to treat these strings by
means of the weak-field approximation.

We are going to consider in this paper supermassive strings. They
are strings with $\eta >> 10^{-3}$.

 The  model of the self-interacting potential
for the scalar complex field is provided by the standard
"Mexican hat" potential. Then, the only consistent way to find
the metric describing an infinitely long cosmic string is to
solve the coupled Einstein-scalar-gauge equations.

We consider the simplest case of a local U(1) symmetry. After spontaneous
symmetry breaking, the manifold of degenerate vacua is a circle;
since this manifold is not simply connected, strings can occur.
The simplest scalar gauge field theory leading to the formation
of cosmic strings is the Abelian Higgs model which contains a
U(1) gauge field $A_{\mu}$ and a complex charged scalar
field $\Phi = R \exp{(i \psi)}$
with a coupling constant $e$ and a potential

\begin{equation}
V(\Phi) = \frac{\lambda}{2} ( \Phi \Phi^{*} - \eta^{2} )
\end{equation}
where $\lambda$ and $\eta$ are two coupling constants. (We use units
where $G = c = 1$.)

The Lagrangian for these fields is,

\begin{equation}
L = - \frac{1}{2} \nabla^{\mu}R \nabla_{\mu}R - \frac{1}{2} R^{2}
(\nabla_{\mu}\psi + e A_{\mu}) (\nabla^{\mu}\psi + e A^{\mu}) -
\lambda (R^{2} - \eta^{2})^{2} - \frac{1}{16 \pi} F_{\mu \nu} F^{\mu \nu}.
\end{equation}
where
$F_{\mu \nu} = \nabla_{\mu} A_{\nu} - \nabla_{\nu} A_{\mu}$.

The complex scalar field and the components of
the gauge field take the form

\begin{eqnarray}
R = R(r)  \,\,\, ,  \,\,\,  \psi = n  \phi \nonumber ,\\
A_{\mu} = \frac{1}{e} [ P(r) - n ] \nabla_{\mu} \phi ,
\end{eqnarray}
\noindent where $n$ is the winding number.

The spacetime metric assumed to be static and cylindrically symmetric,
as shown by Shaver [8], is given by,

\begin{equation}
ds^{2} = - dt^{2} + dr^{2} + dz^{2} + \beta(r)^{2} d\phi^{2}
\end{equation}
with the angular coordinate $\phi$ ranges from 0 to 2 $\pi$ as usual.

To get solutions with the metric regularity at $r=0$,
and with a finite energy,
we take the boundary conditions

\begin{eqnarray}
\lim_{r \rightarrow 0}{R(r)} = 0 \,\,\, and \,\,\, \lim_{r \rightarrow 0}
{P(r)} = n  \nonumber \\
\lim_{r \rightarrow \infty}{R(r)} = \eta \,\,\, and
\lim_{r \rightarrow \infty}{P(r)} = 0.
\end{eqnarray}

Linet [9] proved the Bogomol'nyl inequality on the linear mass density
$\mu$ of these solutions in the case $e^{2} \leq 8 \lambda$.
The lower bound $\mu = \pi |n| \eta^{2}$  correspond a
solution with $e^{2} = 8 \lambda$ and the metric has
necessarily the form (4). We are going to consider the case $n = 1$.
On the axis $r=0$ the function $\beta$ has to satisfy the
following condition

\begin{equation}
\lim_{r \rightarrow 0}{\frac{\beta^{2}}{r^{2}}} = 1 .
\end{equation}

In this case the non-vanishing components of $G^{\mu}_{\nu}$ are

\begin{equation}
G_{z}^{z} = G_{t}^{t} = \frac{1}{\beta} \frac{d^{2} \beta}{dr^{2}}.
\end{equation}

The components of the energy-momentum tensor are

\begin{eqnarray}
T_{t}^{t} = T_{z}^{z} = - \frac{1}{2} [ (R')^{2} + \frac{R^{2} P^{2}}
{\beta^{2}} + 2 \lambda (R^{2} - \eta^{2})^{2} + \frac{(P')^{2}}
{\beta^{2} e^{2}} ] \\
T_{r}^{r} = \frac{1}{2} [ (R')^{2} - \frac{R^{2} P^{2}}
{\beta^{2}} - 2 \lambda (R^{2} - \eta^{2})^{2} + \frac{(P')^{2}}
{\beta^{2} e^{2}} ]  \\
T_{\phi}^{\phi} = \frac{1}{2} [ -(R')^{2} + \frac{R^{2} P^{2}}
{\beta^{2}} + 2 \lambda (R^{2} - \eta^{2})^{2} + \frac{(P')^{2}}
{\beta^{2} e^{2}} ] .
\end{eqnarray}

When we consider $e^{2} = 8 \lambda$ , the function $R$ and $P$ must
obey the following Bogomol'nyi equations

\begin{eqnarray}
P' = \frac{\beta e^{2}}{2} (R^{2} - \eta^{2})  \\
R' = \frac{R P}{\beta}
\end{eqnarray}

and the Einstein equations reduce to,

\begin{equation}
\frac{\beta''}{\beta} = - 8 \pi ( \frac{R^{2} P^{2}}{\beta^{2}} +
\frac{e^{2}}{4} ( R^{2} - \eta^{2} ) )
\end{equation}

Using (11) and (12) in (13), we can evaluate the integration,
obtaining the following equation

\begin{equation}
\beta' = - 4 \pi (R^{2} - \eta^{2}) P + 1 - 4 \pi \eta^{2}.
\end{equation}

Making the transformation

\begin{eqnarray}
\beta = \frac{2 \sqrt{2} \bar{\beta}}{\eta e} \,\, ,   \\
R  = \eta \bar{R}  \,\,\, ,
\end{eqnarray}
and
\begin{equation}
r = \frac{2 \sqrt{2} \bar{r}}{\eta e}  \,\, ,
\end{equation}
as proposed by Shaver[8] , we see that the equations (11) and (14) become,

\begin{eqnarray}
\bar{P}' = 4 \bar{\beta} (\bar{R}^{2} - 1) \,\,  , \\
\bar{\beta}' = - 4 \pi \eta^{2} ( \bar{R}^{2} - 1 ) \bar{P}
+ 1 - 4 \pi \eta^{2}
\end{eqnarray}
and the equation (12) remains the same.

The set of equations (12), (18-19) with the boundary conditions (5)
describe
the cosmic strings. However, the exact solutions of the set of
equations are so far not obtained. Linet [10] attempted to
obtain exact solutions for the special case of a supermassive local
string $4 \pi \eta^{2} = 1$.

For a supermassive local string, we consider $4 \pi \eta^{2} = 1$,
then the equation (19 ) reduces to,

\begin{equation}
\bar{\beta}' = - ( \bar{R}^{2} - 1 ) \bar{P} .
\end{equation}

{}From the equations (20) and (18 ) we obtain that,

\begin{equation}
\bar{P} = \sqrt{ 1 - 4 \bar{\beta}^{2} } \,\,\, ,
\end{equation}
and,
\begin{equation}
2 \frac{\bar{R}'}{\bar{R}} = \frac{( \frac{- \bar{\beta}'}{\bar{P}} )'}
{1 - \frac{\bar{\beta}'}{\bar{P}}}
\end{equation}

Substituting (21) and (22) in (12) , we construct the followying
differential equations for $\bar{\beta}$

\begin{equation}
4 \bar{\beta}^{2} (\bar{\beta}')^{2} + ( 1 - 4 \bar{\beta}^{2} )
\bar{\beta} \bar{\beta}''
+ 2 ( 1 - 4 \bar{\beta}^{2} )^{2} - 2 ( 1 - 4 \bar{\beta}^{2} )^{\frac{3}{2}}
\bar{\beta}' = 0 .
\end{equation}

An exact solution of the differential equation is given by,

\begin{equation}
\bar{\beta} = \frac{1}{2} \sin{( 2 \bar{r})} .
\end{equation}

Then we have ,

\begin{equation}
\bar{P} = \cos{(2 \bar{r})} .
\end{equation}

Substituting (16) and (17) in (24) and (25), we have

\begin{eqnarray}
\beta = \frac{1}{a} \sin{(a r)} \,\, , \\
P = \cos{( a r)} \,\,\, ,
\end{eqnarray}
where $a = \frac{ e }{ 8 \sqrt{2 \pi} }$ .

We conclude that, in this case with a constant energy momentum tensor
as the source,
$R = 0$, what
leads us to consider $\phi = 0$ . Hiscok[5] obtained the same solution.

Recently Novello and Da Silva  [11] obtained the exact solution in
this form for the gravitational field of string related to
the fundamental states of bosons of arbitrary mass.

To obtain the solution in the exterior region of the string we put
$T_{\mu}^{\nu} = 0$. From (8-10) one obtains $R = \eta$ and $P = 0$.

The solution of the equation (13) is then given by $\beta = b r + c$
, where $b$ and $c$ are constants. The constant $c$ can be removed by
shifting the origin, the metric then takes the form

\begin{equation}
ds^{2} = - dt^{2} + dr^{2} + dz^{2} + b^{2} r^{2} d\phi^{2}  .
\end{equation}

In this metric, that is Minkowski flat, the angular coordinate $\phi$
ranges from $0$ to $2 \pi b$. Applying the usual
Darmois-Lichnerowicz junction conditions
(in this case reduces to $g_{\phi \phi}^{+} = g_{\phi \phi}^{-}$ and
$\frac{dg_{\phi \phi}^{+}}{dr} = \frac{dg_{\phi \phi}^{-}}{dr}$,
where + represent the exterior metric and - the interior) at the
boundary $r = r_{0}$, we find that

\begin{equation}
\frac{1}{a} \sin{(a r_{0})} = b r_{0}  \,\, ,
\end{equation}
and,
\begin{equation}
\cos{(a r_{0})} = b .
\end{equation}

The same results can be obtained from the junction conditions used by
Hiscock [5]: $g_{\mu \nu}^{+} = g_{\mu \nu}^{-}$ and
$K_{\mu \nu}^{+} = K_{\mu \nu}^{-}$, where $K_{\mu \nu}$ is
the extrinsic curvature.

Evaluating the linear energy density we obtain,

\begin{eqnarray}
\mu = \int_{0}^{r_{0}} \,\, \int_{0}^{2 \pi} \,\, T_{t}^{t}
\beta d\phi dr \nonumber \\
\,\,\, = \frac{1}{4} [1 - \cos{(a r_{0})}].
\end{eqnarray}

{}From the equations (30) and (31), we get

\begin{equation}
b = (1 - 4 \mu)
\end{equation}

The exterior metric takes the form

\begin{equation}
ds^{2} = - dt^{2} + dz^{2} + dr^{2} + (1 - 4 \mu)^{2} r^{2} d\phi^{2}
{}.
\end{equation}

Here we can define an angular coordinate $\phi' = (1 - 4 \mu) \phi$ that
ranges from $0$ to $2 \pi (1 - 4 \mu)$, giving a conical
angular deficit $\delta \phi = 8 \pi \mu$.

{}From (8-10) one finds that,

\begin{equation}
T_{t}^{t} (\phi =0) = T_{t}^{t} (A_{\phi}) = 2 \lambda \eta^{4}
\end{equation}
in the interior region of the string.

Since the $\phi$-field acquires the null value in the string core as
shown by Linde[12] during the inflationary period we may
consider that after the phase transition the $\phi$-field settles
down to the null value producing constant energy density.
The gauge field also contributes the same amount of constant
energy density.

However it should be remarked that this configuration of the
string is possible only in the case of a supermassive local strings
with $4 \pi \eta^{2} = 1$.

The authors would like to thank CNPq and UENF for a grant.

\bigskip
\bigskip
\bigskip

\centerline{\bf{References}}

\bigskip
\bigskip

[1] D.A. Kirzhnits, JETP Lett. \b{15}, 745 (1972). \par
[2] A.D. Linde, Rep. Prog. Phys. \b{42}, 389 (1979). \par
[3] T.W. Kibble, J. Phys. \b{A9}, 1387 (1976). \par
[4] A. Vilenkin, Phys. Rev. \b{D23}, 852 (1981). \par
[5] W. Hiscock, Phys. Rev. \b{D31}, 3288 (1985). \par
[6] J. R. Gott III, Astrophys. J. \b{288}, 422 (1985). \par
[7] A. K. Raychauduri, Phys. Rev. \b{D41}, 3041 (1990).  \par
[8] E. Shaver, GRG \b{24}, 187 (1992). \par
[9] E. B. Linet, Phys. Lett. \b{124A}, 240 (1990). \par
[10] E. B. Linet, Class. Quantum Grav. \b{7}, L75 (1990). \par
[11] M. Novello and M.C. Motta da Silva, Phys. Rev. \b{D48}, 5017
(1993). \par
[12] A.D. Linde, Phys. Rev. \b{D50}, 2456 (1994).

\end{document}